\documentclass[aps,amsmath,amssymb,prb,twocolumn,showpacs,superscriptaddress,floatfix]{revtex4-1}
\usepackage{bm}
\usepackage{epsfig}
\usepackage{graphicx}
\usepackage[hypertex]{hyperref}

\newcommand{\TI}{\scriptscriptstyle{\mathrm{TI}}}

\DeclareMathOperator{\h}{\check{\mathcal H}}
\DeclareMathOperator{\U}{\check{\mathcal U}}
\DeclareMathOperator{\Uu}{\hat{\mathcal U}}

\DeclareMathOperator{\tUu}{\hat{\tilde {\cal U}}}
\DeclareMathOperator{\hbdg}{{\mathcal H}_{\scriptscriptstyle{\mathrm{BDG}}}}
\begin{document}

\title{Instability of the topological order and localization of the edge states  in HgTe quantum wells coupled to s-wave superconductor}
%Interplay of superconductivity and topological order}
\author{I.\,M.\,Khaymovich }
\affiliation{Argonne National Laboratory, Argonne, IL 60439, USA}
\affiliation{ Institute for Physics of Microstructures, Russian Academy of Sciences, 603950 Nizhny Novgorod, GSP-105, Russia}
\author{N.\,M.\,Chtchelkatchev }
\affiliation{Argonne National Laboratory, Argonne, IL 60439, USA}
\affiliation{Institute for High Pressure Physics, Russian Academy of Sciences, Troitsk 142190, Moscow Region, Russia}
\affiliation{L.\,D.\,Landau Institute for Theoretical Physics,
Russian Academy of Sciences, 117940 Moscow, Russia}
\affiliation{Department of Theoretical Physics, Moscow Institute
of Physics and Technology, 141700 Moscow, Russia}
\author{V.\,M.\,Vinokur }
\affiliation{Argonne National Laboratory, Argonne, IL 60439, USA}

\date{\today}

\begin{abstract}
Using the microscopic tight-binding equations we derive the effective Hamiltonian for the two-layer hybrid structure comprised of the
two-dimensional HgTe quantum well-based topological insulator (TI) coupled to the s-wave isotropic superconductor (SC)
and show that it contains terms describing mixing of the TI subband branches by the superconducting correlations induced by the proximity effect.
We find that the proximity effect breaks down the rotational symmetry of the TI spectrum.
We show that the edge states not only acquire the gap, as follows from the standard theory, but can also become localized by the
Andreev-backscattering mechanism in a small coupling regime.  In a strong coupling regime the edge states merge with the bulk
states, and the TI transforms into an anisotropic narrow-gap semiconductor.
\end{abstract}

\maketitle

\section{Introduction}
A topological insulator (TI), a material in which the electronic spectrum possesses an energy gap in the bulk but has the
special, so-called topologically protected, edge (surface) states falling into this gap, is one of the focal points of current condensed matter
studies.~\cite{Volkov_TI,Zgang_rev,rev1,rev2} Topological insulators hold high technological promise since due to the ability of their topologically protected
edge states to carry nearly dissipationless current they can be utilized in integrated circuits.  Thus the question how robust the edge states are
with respect to hybridization with the electronic states in the leads has become one of the focal points of current  TI-related research.

Three dimensional (3D) TI have the Dirac spectrum with the finite mass in the bulk while the spectrum of the surface states is
 massless.\cite{3DTh,3DTh1,3Dexp,3Dexp0,3Dexp1,3Dexp2,Zgang_rev} Two-dimensional (2D) TI has the gapless helical edge states.\cite{Volkov_TI,kane-mele,BHZ}
 The surface states in 3D TI and the edge states in 2D TI are topologically protected and are robust against all time-reversal-invariant local
  perturbations.\cite{Zgang_rev} It was shown experimentally that 2D TI-state appears in HgTe/CdTe quantum wells (QW).\cite{Konig,2Dexp} The TI in three-dimensional (3D) materials was found in Bi$_{1-x}$Sb$_x$, Bi$_2$Se$_3$ and so on.\cite{3Dexp,3Dexp0,3Dexp1,3Dexp2}

\begin{figure}[h]
  \includegraphics[width=0.9\columnwidth]{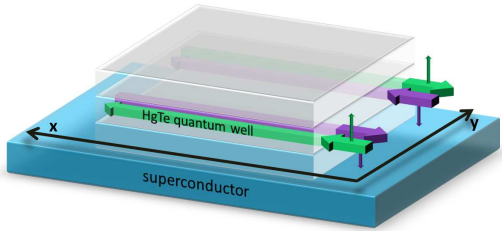}\\
  \caption{(Color online) Sketch of the 2D topological insulator coupled to a superconductor. The thick arrows show schematically the edge states. }\label{fig2}
\end{figure}

Of special interest are the states that develop at the interface between a TI and an s-wave superconductor (SC), like, e.g., in Fig.~~\ref{fig2}, where the proximity effect generates a superconducting pairing. it was shown that at the interface of 3D TI coupled to s-wave superconductor $p_x+ip_y$-superconducting state appears but without time-reversal symmetry breaking.\cite{Fu} Numerical calculations showed that the proximity of the superconductor leads to a significant renormalization of the original parameters of the effective model describing the surface states of a topological insulator.\cite{stanesku} For 2D topological insulator coupled to s-wave superconductor it is known that the momentum-independent gap enters in the edge spectrum.\cite{gang,gang1,gang2,Jiang,Jiang1}

Two-dimensional TIs  have a unique property: their parameters can be tuned over wide ranges of their values by the appropriate choice of the
QW width $d$.\cite{Konig} In particular, the main parameter of 2D TI, the gap in the bulk spectrum, $M$, can be changed from zero up to room
temperature energy scale. Phenomenological treatment of the proximity effect is based on the assumption that the gap $M$ in the bulk spectrum of TI is
much larger than the characteristics energy of the induced superconducting correlations. Unfortunately numerical calculations do not give the answer
how the spectrum of the TI-SC system develops in this regime.\cite{Jiang,Jiang1} In what follows we will focus on the gaps $M$ of the order of the
energy of superconducting correlations.

In this Paper we investigate the effect of superconducting  correlations on the topologically protected edge states and the bulk spectrum in the
2D topological insulator brought into a contact with the SC layer, see Fig.~\ref{fig2}, and show the emergence of the SC correlations in the TI
similarly to what was observed in GaAs containing a two-dimensional electron gas.\cite{Takayanagi,Batov}
Ordinarily, the effective Hamiltonian of the TI in question is constructed from the symmetry considerations, see e.g., Ref.~\onlinecite{gang,gang1,gang2},
and has the trivial structure of the induced superconducting potentials. We demonstrate that the symmetry reasons suggest the additional non-diagonal
(in the subband space) terms in the Hamiltonian. Moreover, using the microscopic tight-binding equations we derive analytically
the additional terms in the effective Hamiltonian of the TI describing coupling to the $s$-wave isotropic superconductor (SC) placed on top of it.
These terms become especially important in the case where the bare gap parameter $M$ of the TI
becomes comparable to the characteristic energy of the induced superconducting correlations.
We show, further, that the interplay of the superconducting and ``topological'' interactions is essential and results in several effects,
in particular, the collapse of the topological order in TI.
We find that while ``topologicaly protected'' edge states can ensure undisturbed
propagation of the charge (spin) carriers, superconducting correlations can block the edge current causing a peculiar localization effect.

On the qualitative level our results can be summarized as follows: in the absence of superconducting correlations the edge states spectrum
is linear near the Fermi surface, comprising of two
counter-propagating  electron and two hole branches, respectively.\cite{Zgang_rev} Importantly, the edge spectrum is
isotropic in a sense that it does not depend on the orientation of
the edges with respect to crystallographic axes, although the edge
electronic states wave functions are orientation dependent.
Namely, there is a phase difference between the wave function
components corresponding to (E) or (H)-subbands in TI. The bulk
spectrum is characterized by the gap $M$. At small SC-TI coupling
(the quantitative criteria will be given below) the edge-states
spectrum acquire a gap $E_{\mathrm g}$. Furthermore,
superconducting correlations mix the subband branches and thus
turn the resulting electronic spectrum in TI (edge and bulk),
 anisotropic, and $E_{\mathrm g}$ starts to depend on orientation of the edge.
This can cause localization of the low-lying edge states since an
inevitable bending of the edge will create the regions along the
edge were the edge-particle energy $\varepsilon<E_{\mathrm g}$
thus getting locked between the turning points where
$\varepsilon=E_{\mathrm g}$. At the turning points electron and
hole excitations undergo Andreev reflection and form the localized
Andreev bound edge states, see Fig.~\ref{fig1}a. As the
characteristic energy of the induced pairing amplitude exceeds $M$
then the gap of the continuum bulk states of TI collapses and TI
behaves like the highly anisotropic narrow-gap semiconductor.

\section{Effective Hamiltonian}

The low energy Hamiltonian of the two-dimensional (2D) topological
insulator formed in the HgTe QW has the form:\cite{BHZ,Zgang_rev}
\begin{gather}\label{BHZ_Ham}
   \h=\begin{pmatrix}
       \hat H & 0 \\
       0 & \hat{\tilde H} \\
     \end{pmatrix},
\end{gather}
where $\hat H=\epsilon_k+d_i\hat\sigma^i$, $i=\{1,2,3\}$; $\hat\sigma^i$ are the Pauli matrices acting in the subband (isospin) space; $\epsilon_k=C-D k^2$. We choose the frame of reference so that $\vec d=( k_x A,- k_y A, M-B k^2)$. Here  $A$, $B$, $C$, $D$ and $M$ are material parameters. The lower block of the Hamiltonian, $\hat{\tilde H}=\hat\rho^T \hat H^* \hat\rho = \epsilon_k-d_i(-k)\hat\sigma^i$, where $\hat\rho=i\hat\sigma_y$ is the metric tensor in the spinor space. The chosen representation for $\h$ enables us to employ the machinery of the tensor bispinor algebra developed for Dirac Hamiltonian.

%.
\begin{figure}[t]
  \includegraphics[width=0.95\columnwidth]{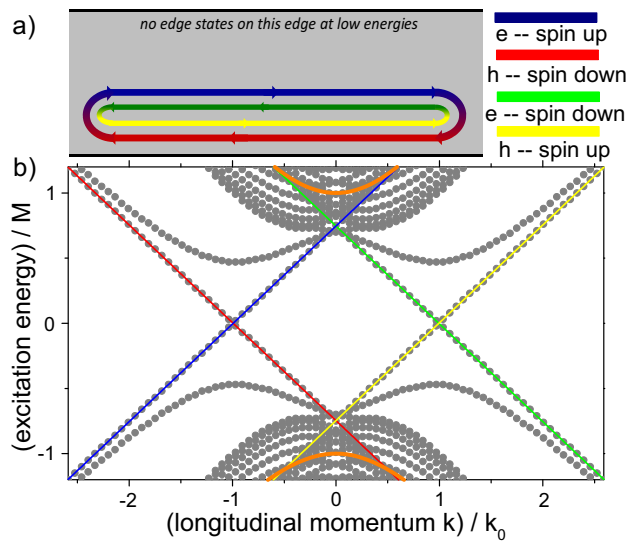}\\
  \caption{(Color online)
  a) A sketch of localized Andreev edge states in TI.
  b) Excitation spectrum in the 2D TI
  with proximity
  induced superconducting correlations: the edge modes at one TI-edge acquired the gap while the edge modes on the opposite edge remain gapless.
  Solid lines show the edge states and bulk spectrum boundaries in TI without superconducting correlations.
  The parameters are chosen as: $mk_Ft_a^2/(2\pi\hbar^2M)=1$, $t_b=\sqrt{|B_-|/|B_+|}t_a$ and
the orientation angle $\varphi=0$. }\label{fig1}
\end{figure}
To construct the convenient form of the Bogoliubov -- de Gennes (BdG) Hamiltonian describing superconductivity, we introduce
the time reversal symmetry operator,
\begin{gather}
\check{\mathcal{T}} = \begin{pmatrix}
0 & 0 & 0 & -1\\
0 & 0 & 1 & 0\\
0 & -1 & 0 & 0\\
1 & 0 & 0 & 0\\
\end{pmatrix}{\mathcal{C}} = -\hat\tau_1\otimes i\hat\sigma_2\, \mathcal{C}\,,
\end{gather}
where ${\mathcal{C}}$ is the operator of the complex conjugation and $\hat\tau_i$,
are the Pauli matrices acting in spin space.
Then the time-reverse of the BHZ-Hamiltonian \eqref{BHZ_Ham} is $\check{\mathcal{T}}\h\check{\mathcal{T}}^{-1} = \h$.
The BdG Hamiltonian is
\begin{gather}\label{BdG-Ham}
   \hbdg=\begin{pmatrix}
       \h + \U & \check\Delta_{\rm\scriptscriptstyle TI} \\
       \check\Delta^+_{\rm\scriptscriptstyle TI} & -\h-\check{\mathcal{T}}\U\check{\mathcal{T}}^{-1} \\
     \end{pmatrix}\,,
\end{gather}
where $\check\Delta_{\rm\scriptscriptstyle TI}$ is the effective proximity induced superconducting pairing matrix coupling the spin and subband spaces.
The effective chemical potential shift appearing in the BCS theory has matrix form $\U$. There is no reason to believe that the matrix structure of $\U$ and $\check\Delta_{\rm\scriptscriptstyle TI}$ is necessary trivial like, e.g., in Ref.~\onlinecite{gang,gang1,gang2}; symmetry considerations in fact allow nontrivial shape of theses matrices. So both, $\check\Delta_{\rm\scriptscriptstyle TI}$ and $\U$ will be found below microscopically.

To proceed further we present the Hamiltonian describing the TI-SC coupling in a form:
\begin{equation}\label{tunn_Ham}
 H =  H_{sc}+ H_{2D}+ H_{\rm int} \,.
\end{equation}
The superconducting part is
\begin{multline}
 H_{sc}= \sum\limits_{s=\uparrow,\downarrow}\int d^3r\Psi^+_s
(\textbf{r})\left(\epsilon_{sc}-\mu\right)\Psi_s
(\textbf{r}) +
\\
 \int d^3r\left(\Delta\Psi^+_\uparrow
(\textbf{r})\Psi^+_\downarrow
(\textbf{r})+\Delta^*\Psi_\downarrow
(\textbf{r})\Psi_\uparrow (\textbf{r})\right)
\end{multline}
where,
$\Psi_{\uparrow(\downarrow)}$ ($\Psi^+_{\uparrow(\downarrow)}$) are the field annihilation (creation)
operators for the state with the spin up (down),
$\Delta$ is the superconducting gap, $\epsilon_{sc}$ is the single electron kinetic energy, and $\mu$ is the Fermi energy.

\begin{figure}[t]
  \includegraphics[width=0.8\columnwidth]{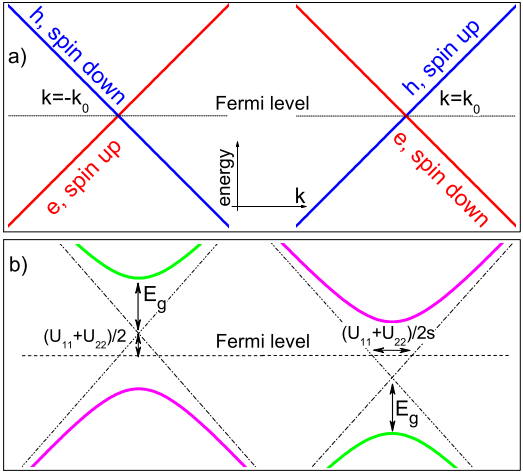}\\
  \caption{(Color online) The edge states at the TI boundary for a weak coupling where the edge states become gapped. a) Unperturbed edge-spectrum.
  b) Edge states with the proximity-induced gap.  The shift of the zero point reflects the difference in the original chemical potentials.
  }\label{fig6}
\end{figure}
The second quantization representation for the TI Hamiltonian is written in the basis of the Wannier functions for particles with spin $s$:
\begin{multline}
 H_{2D,s}= \sum\limits_{\textbf{R}\textbf{R}^\prime,s}\sum\limits_{\sigma,\sigma'=a,b}
c^+_{s\textbf{R},\sigma}\left(\epsilon_{2D,s}(\textbf{R}\sigma,\textbf{R}^\prime\sigma')+
\right.\\\left.
C\delta(\textbf{R},\textbf{R}^\prime)\delta_{\sigma,\sigma''}\right)
c_{s\textbf{R}^\prime,\sigma'}
\end{multline}
where $\epsilon_{2D,s}(\textbf{R},\textbf{R}^\prime)$ is the lattice representation of the BHZ-model\cite{Jiang,Jiang1} \eqref{BHZ_Ham} (see Appendix~\ref{ap}). Then $ H_{2D}=\sum_{s} H_{2D,s}$.

Finally, $ H_{\rm int}$ reflects the electronic tunneling between the SC and TI:
\begin{equation}\nonumber
 H_{\rm int}=\sum\limits_{\textbf{R},s}\sum\limits_{\sigma=a,b}\left(t_{\sigma,\mathbf{R}} \Psi^+_s(\mathbf{R}) c_{s\mathbf{R},\sigma}+t_{\sigma,\mathbf{R}}^* c^+_{s\mathbf{R},\sigma} \Psi_s(\mathbf{R})\right)\,,
\end{equation}
where $ c_{\uparrow(\downarrow)\textbf{R},a}$ is the superposition
of $\left|\Gamma_6,\pm\tfrac{1}{2} \right>$,
$\left|\Gamma_8,\pm\tfrac{1}{2} \right>$ and $
c_{\uparrow(\downarrow)\textbf{R},b}$ refers to the subband
$\left|\Gamma_6,\pm\tfrac{3}{2} \right>$. Integrating out the bulk
superconductor variables $ \Psi_s(\mathbf{R})$ using the method,
developed in Ref.~\onlinecite{Volkov,Meln_Kopnin_proxim_Delta}, one obtains
the effective BdG-Hamiltonian \eqref{BdG-Ham} for the homogeneous
tunneling amplitudes $t_{\sigma\textbf{R}} = t_{\sigma} $, with
the matrix superconducting order parameter and the effective
chemical potential shift having the form:
\begin{gather}
\check\Delta_{\rm\scriptscriptstyle TI} = \begin{pmatrix}
\hat\Delta_{\rm\scriptscriptstyle TI} & 0 \\
0 & \hat{\tilde\Delta}_{\rm\scriptscriptstyle TI}\\
\end{pmatrix} \ ,\qquad \U = \begin{pmatrix}
\Uu & 0 \\
0 & \tUu\\
\end{pmatrix} \,.
\end{gather}
Here
\begin{gather}\hat\Delta_{\rm\scriptscriptstyle TI} = -\frac{m k_F}{2\pi\hbar^2}\begin{pmatrix}
t_a^{*2} & t_a^*t_b^* \\
t_a^*t_b^* & t_b^{*2}\\
\end{pmatrix} \ , \qquad \hat{\tilde \Delta}_{\rm\scriptscriptstyle TI}=\hat\rho^T \hat\Delta_{\rm\scriptscriptstyle TI} \hat\rho\,,
\\\label{eqU}
\Uu = \frac{m}{2\pi\hbar^2a_{\rm\scriptscriptstyle TI}}\begin{pmatrix}
|t_a|^2 & t_a^*t_b \\
t_a t_b^* & |t_b|^2\\
\end{pmatrix}\,,  \qquad {\tUu}=\hat\rho^T \Uu \hat\rho \,,
\end{gather}
$m$ and $k_F$ are the effective mass and the Fermi momentum of the
bulk superconductor respectively, $a_{\rm\scriptscriptstyle TI}$
is the characteristic length scale of the order of  the lattice
constant in TI. Since $\Delta_{\TI}\ll\Delta$, the proximity
induced parameters are independent of
$\Delta$.~\cite{Meln_Kopnin_proxim_Delta} In Ref.~\onlinecite{gang,gang1,gang2} the potentials $\hat\Delta_{\rm\scriptscriptstyle TI}$ and $\Uu$ were diagonal (trivial) while the offdiagonal terms were missed.

\begin{figure}[t]
                      \includegraphics[width=0.95\columnwidth]{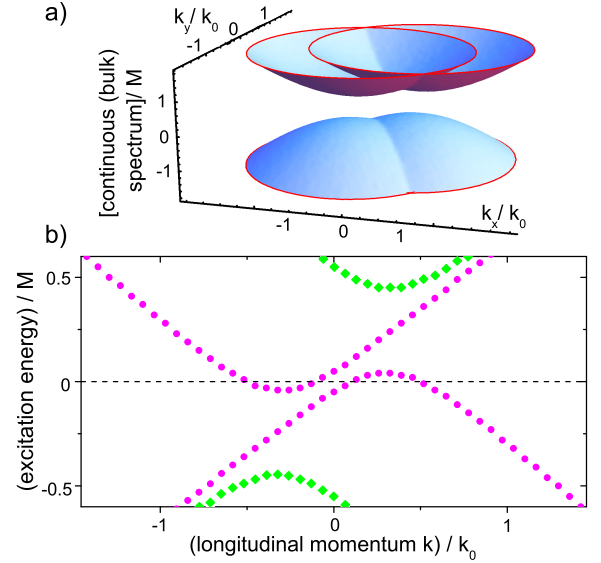}\\
                 \caption{ (Color online) Bulk and edge states for the intermediate coupling. a) Energy of the bulk states as function of $k_x, k_y$. Without superconductivity, the bulk states dispersion is isotropic, $E(k)=\epsilon(k)\pm\sqrt{A^2 k^2+(M-Bk^2)^2}$,\cite{Zgang_rev} where $k=\sqrt{k_x^2+k_y^2}$. Superconducting correlations make it anisotropic as follows from the noncommutativity of  $\check\Delta$ and/or $\U$ with $\h$ in Eq.\eqref{BdG-Ham}.
                 b) Gapped edges states. The colors for the families of the dispersion curves are the same as in Fig.~\ref{fig6}.  The parameters
                 are chosen as: $mk_F/(2\pi\hbar^2) t_a^2/M=1$, $t_b=\sqrt{|B_-|/|B_+|}t_a\exp(i\pi/6)$ and the orientation angle $\varphi=\pi/2$.  }\label{fig6a}
        \end{figure}

For numerical calculations we take typical parameters:
$A=3.8$~eV{\AA}, $B=-56.2$~eV{\AA}$^2$, $D=-38.7$~eV{\AA}$^2$.
Without a loss of generality we take the energy-shift parameter
$C=0$. We do not fix $M$ ($-10$meV$\lesssim M<0$) and use it as
the energy unit. Our numerical and analytical calculations show
the approximate symmetry relation that satisfies the spectrum of
$\hbdg$: $\chi
E(\mathbf{k}/\chi,M/\chi,t_a/\sqrt\chi,t_b/\sqrt\chi)\approx
E(\mathbf{k},M,t_a,t_b)$, where $\chi$ is a dimensionless scaling
parameter. The scaling relation appears since $M$ is much smaller
than the energy scales one can construct form $A$, $B$ and $D$. In
addition, $M$ appears to be the most sensitive to the HgTe layer
width: it changes with it by several orders of magnitude while the
other parameters change by $\sim20\%$ and their changes very
slightly modify the spectrum.\cite{Zgang_rev}

 \begin{figure}[t]
             \includegraphics[width=0.95\columnwidth]{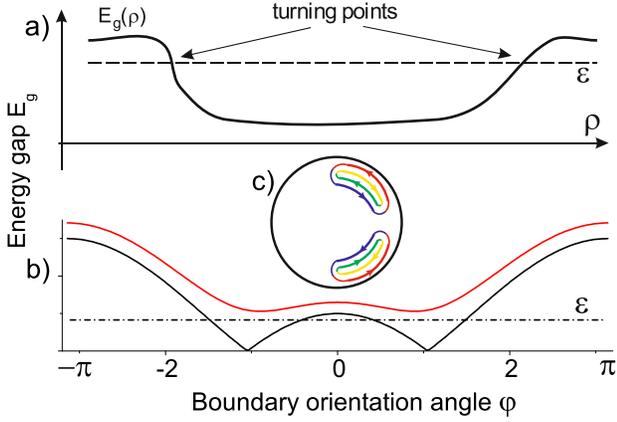}\\
            \caption{ (Color online) Energy landscape for the edge states.  a) Sketch of the energy landscape along the edge (parametrized by the coordinate $\rho$).
            The magnitude of the gap may change as a result of spatial fluctuations (like change in a shape of the TI boundary or fluctuations in tunneling amplitudes)
            and the state with the energy
             $\varepsilon$ would appear trapped between the turning points where $\varepsilon<E_g$ forming
             Andreev bound edge state like it is shown in           Fig.~\ref{fig1}.
  %
  %
  \iffalse
   .  There are points on the boundary where $E_g$ changes as the function of the coordinate $\rho$ along the boundary due to bending of TI-boundary or change of tunnel matrix elements $t_a$ and $t_b$. At these ``turning points'' electron and hole edge states (going in the opposite direction) with the energy $\varepsilon<E_g$ undergo Andreev reflection and may form the Andreev bound edge state like it is shown in Fig.~\ref{fig1}.
   \fi
%
  %
            b) Calculated $E_g$ as function of the edge orientation angle $\varphi$ for the ratio $t_a/t_b\sqrt{|B_-/B_+|}\exp(-i\pi/3)$ equal to 1 (upper curve)
            and 2 (bottom curve). Such $E_g$ behavior can be observed in the sample shaped into a disc. c) Localized edge states in the sample shaped into a disc for energy $\epsilon$ corresponding to the dash-dot line in Fig.~\ref{fig7}b.}\label{fig7}
        \end{figure}
First we discuss ``weak'' superconductivity where superconducting
correlations induced in TI can be treated perturbatively. In this
case matrix elements of $\check{\Delta}_{\rm\scriptscriptstyle
TI}$ and $\U$ are smaller than the gap in the continuum spectrum,
$M$, in the bulk of TI. In the absence of superconducting
correlations $\check{\Delta}_{\rm\scriptscriptstyle TI}=0$ and
$\U=0$, and there are two electron and two hole edge states at
each TI surface.  The edge states have the linear dispersion law
with the velocity $s=A|\sqrt{B_+B_-}/B|$, where $B_\pm = B \pm D$.
They cross the Fermi energy at $k=k_0=DM/(A\sqrt{B_+B_-})$ and
$k=-k_0$, see Fig.~\ref{fig6}a. We denote the wave functions of
the electron and hole edge states near $k=k_0$ as
$\psi^{(1)}=(\psi_{\rm edge},\hat 0,\hat 0,\hat 0)^\tau$ and
$\psi^{(2)}=(\hat 0,\hat 0,\psi_{\rm edge},\hat 0)^\tau$,
respectively, where $\hat 0$ is the zero spinor in the subband
space,
    \begin{multline}\label{Insulating_edge_function}
                  \psi_{\rm edge} = \frac{e^{-i\hat\sigma_z\varphi/2}}{\sqrt{2|B|}}\left(\sqrt{|B_-|}\atop-\sqrt{|B_+|}\right)\times
\\
                  \left(e^{-\lambda_+ \mathbf r\cdot \mathbf{n}}-e^{-\lambda_- \mathbf r\cdot \mathbf{n}}\right)e^{ik \mathbf{r}\cdot\mathbf{l}}\,,
     \end{multline}
$k$ is the momentum component parallel to the edge, $\mathbf r=(x,y)$, $\mathbf l$, and $\mathbf n$ are the unit vectors directed along the TI
boundary and perpendicular to it correspondingly [$\mathbf{l}\times \mathbf{n}$ is aligned with the $OZ$ axis], and $\varphi$ is the angle between
$\mathbf l$ and $OX$ axis. The decay length scales of the edge states into the bulk of the topological insulators are:
          \begin{gather}\label{Insulating_edge_lambda}
                      \lambda_{\pm} = \lambda_0 \pm \sqrt{\left(k - \frac{D}{B}\lambda_0\right)^2 + \frac{A^2}{4B^2}-\frac{M}{B}},
           \end{gather}
where $\lambda_0 = A/(2\sqrt{B_+B_-})$. We stress that spinor
components of $\psi_{\rm edge}$ depend on the TI-boundary  orientation.

\begin{figure}[th]
  \includegraphics[width=0.9\columnwidth]{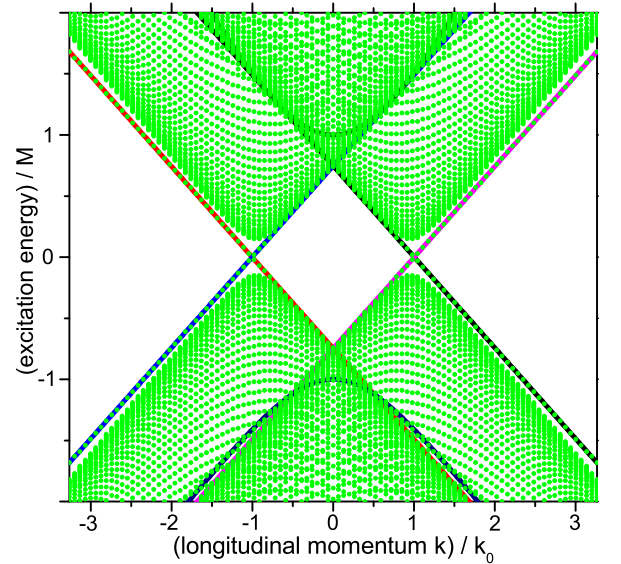}\hspace{10mm} \\
  \caption{(Color online) Excitation spectrum in 2D TI for the proximity induced potentials being of
  the same order as the gap in TI in the absence of a SC.  The gap between the branches of the continuum spectrum collapses and
  TI acquires metallic conductivity with the relativistic spectrum similar to that in graphene.
  Solid lines correspond to the edge states and the bulk spectrum boundaries in TI
   without superconducting correlations. Parameters are chosen as:
   $mkF/(2\pi\hbar^2) t_a^2/M=12$, $\varphi=0$, and $t_b=\sqrt{|B_-|/|B_+|}t_a$.}\label{fig3}
\end{figure}
The dispersion law of the edge states near $k=k_0$ within the perturbation theory taking in the account
the superconducting correlations acquires the form:
    \begin{gather}%\label{}
        \epsilon_{1,2}(k)=(\mathcal{U}_{11}+\mathcal{U}_{22}\pm\omega(k))/2,
    \end{gather}
where $\omega=\sqrt{(2s (k-k_0)+\mathcal{U}_{11}-\mathcal{U}_{22})^2+4E_g^2}$; $\mathcal{U}_{ii}$, $i=1,2$ are the matrix elements of $\U$
with respect to the states $\psi^{(1,2)}$ and $E_g=|(\check{\Delta}_{\rm\scriptscriptstyle TI})_{12}|$.
They can be parameterized through $\mathcal T_\pm = \left|t_a\sqrt{|B_-|}-t_b\sqrt{|B_+|}e^{\pm i\varphi}\right|$. So,
       \begin{gather}
                  E_g=\mathcal T_+\mathcal T_-,
        \end{gather}
the matrix elements in the see Eq.\eqref{eqU} become $\U_{11(22)}=\alpha \mathcal T_{+(-)}^2$, and
 $\alpha = (k_F
a_{\rm\scriptscriptstyle TI})^{-1}$.

One now sees that the spectrum of the edge states becomes dependent upon the orientation of the boundary orientation with respect to
crystallographic axis. This resembles the spectrum that often appear in the 3D TI which that are referred to as ``strong'' TI.\cite{stanesku}

%This is not a miracle. It is known that, e.g., 3D TI can be anisotropic and then it is referred to ``strong'' TI.\cite{stanesku}}

There is a wealth of the possible coupling-induced behaviours of the edge states energy spectrum.
If $\mathcal T_+=0$ or $\mathcal T_-=0$, then $E_g=0$ as well; the situation where $E_g$ is very small is also common.
A particular picture  depends on the edge orientation angle $\varphi$ and/or on the tunneling amplitudes, $t_a$ and $t_b$.
Shown in the Fig.~\ref{fig1} is the
situation where at one boundary of the TI-strip the edge states remain gapless ($E_g=0$) while at the opposite boundary $E_g\neq0$
and the edge states have the gap. The stripe within which the edge states are confined has a finite length (see in Fig.~\ref{fig1}),
there are points at the edge where the TI-boundary changes its direction and,
at the same time, the value of $E_g$ changes. At these ``turning points'' electron and hole edge (going in the opposite direction) states with
the energy smaller
than $E_g$ undergo the Andreev reflection and form the \textit{bound} Andreev edge state, see Figs.~\ref{fig1},\ref{fig7}.
Illustrated in Figs.~\ref{fig6}-\ref{fig6a} is the structure of the the edge state energy levels in the case where $E_g$ is finite.

\iffalse
What happens with the edge state energy levels when $E_g$ is finite is illustrated in Fig.~\ref{fig6}-\ref{fig6a}.
However, the level splitting is not a general situation. It is possible that $E_g$ is very small or even equal to zero when
$\mathcal T_+=0$ or $\mathcal T_-=0$.
A particular picture  depends on the edge orientation angle $\varphi$ and/or on the tunneling amplitudes, $t_a$ and $t_b$. Shown in the Fig.~\ref{fig1} is the
situation where at one boundary of the TI-strip the edge states remain gapless ($E_g=0$) while at the opposite boundary $E_g\neq0$
and the edge states have the gap. The strip in Fig.~\ref{fig1} is not infinite, there are points at the edge where the TI-boundary changes its direction and,
at the same time, the value of $E_g$ changes. At these ``turning points'' electron and hole edge (going in the opposite direction) with the energy smaller
than new $E_g$ undergo Andreev reflection and form the Andreev bound edge state, see Figs.~\ref{fig1},\ref{fig7}.
\fi

Now we discuss a general nonperturbative situation. The excitation spectrum in 2D TI accounting for the proximity
induced superconducting correlations in the case where
proximity induced potentials in TI are of the same order as the gap in TI without superconductor on top is shown in Fig.~\ref{fig3}.
The gap between the branches of the continuum spectrum nearly closes and TI acquires effectively metallic
conductivity with the relativistic spectrum similar to that
in graphene. Solid lines correspond to the edge states and bulk spectrum boundaries in TI without superconducting correlations.

\section{Conclusions}
To conclude, we investigated topologically protected edge states
in QW of HgTe sandwiched between CdTe and demonstrated that the
$s$-wave isotropic superconductor placed on top of CdTe layer
induces superconducting correlations in the TI revealing the built
in anisotropy of TI which did not affect the spectrum when
superconducting correlations were absent. The form of the edge
states spectrum essentially depends on the edge orientation with
respect to crystallographic directions of the TI. Depending on the
coupling between the superconductor and 2D TI, different scenarios
can be realized: (i) the edge states of the topological insulator
acquire a gap,
 (ii) the edge states hybridize into the Andreev localized edge state and/or  (iii) the gap separating the continuum and the edge modes collapses and
 TI becomes the narrowgap (anisotropic) semiconductor.  Our predictions can be verified by means of, for example, scanning tunnelling spectroscopy measurements
 of the spectra showed in Fig.~\ref{fig6} where the shift of the zero point $U_{11}+U_{22}$ can be tuned by the gate placed on top of the CdTe layer.

\textit{Note}: After this work has been completed we became aware of the recent experiments
on InAs/GaSb QW.\cite{2DexpAS} which revealed the 2D TI state.\cite{2DexpAS}  Since this novel TI
is expected to be well described by the BHZ-model, our results apply to InAs/GaSb QW coupled to s-wave superconductor as well.

\section{Acknowledgments}
This work was supported by the U.S. Department of Energy Office of
Science under the Contract No. DE-AC02-06CH11357, the work of IMK
and NMC was partly supported by the Russian president foundation
(mk-7674.2010.2) under the Federal program ``Scientific and
educational personnel of innovative Russia''.

\appendix

\section{Lattice model \label{ap}}
Microscopic description of the coupling between TI and the superconductor developed on the base of the standard lattice
regularization of the BHZ-model~\eqref{BHZ_Ham} replacing its parameters by:\cite{Zgang_rev}
\begin{widetext}\
\begin{gather}
\epsilon_k=C-2Da^{-2}\left[2-\cos k_x a - \cos k_y a\right],
\\
\vec d=\left(Aa^{-1}\sin k_x a,-Aa^{-1}\sin k_y a, M-Ba^{-2}\left[2-\cos k_x a - \cos k_y a\right]\right).
\end{gather}

This corresponds to the quadratic Bravais lattice (with the translation vectors $\textbf{a}_1 = a\textbf{x}_0$, $\textbf{a}_2 = a\textbf{y}_0$)
with two type of states (corresponding annihilation operators are $\hat c_{a\textbf{R}}$ and $\hat c_{b\textbf{R}}$)
on each site replying to subband states.
Therefore the Hamiltonian describing TI in the second quantization representation in the basis of the Wannier functions for particles with spin $s$, takes the form:
\begin{gather}
\hat H_{2D,s}= \sum\limits_{\textbf{R}\textbf{R}^\prime,s}\sum\limits_{\sigma,\sigma'=a,b}\hat
c^+_{s\textbf{R},\sigma}\left(\hat\epsilon_{2D,s}(\textbf{R}\sigma,\textbf{R}^\prime\sigma')+
C\delta(\textbf{R},\textbf{R}^\prime)\delta_{\sigma,\sigma''}\right)\hat
c_{s\textbf{R}^\prime,\sigma'} \ ,
\end{gather}
where
%\begin{subequations}
%\begin{align}
%\hat\epsilon_{2D,\uparrow}(\textbf{R}\tilde\sigma,\textbf{R}^\prime \tilde\sigma) & = \delta_{\textbf{R}\textbf{R}^\prime}\left[\left(M - \frac{4B}{a^2}\right)\tilde\sigma - \frac{4D}{a^2}\right] + \left(\frac{4B}{a^2}\tilde\sigma + \frac{4D}{a^2}\right)\left(\delta_{\textbf{R+a}_1,\textbf{R}^\prime}+\delta_{\textbf{R-a}_1,\textbf{R}^\prime}+\delta_{\textbf{R+a}_2,\textbf{R}^\prime}+\delta_{\textbf{R-a}_2,\textbf{R}^\prime}\right)\\
%\hat\epsilon_{2D,\downarrow}(\textbf{R}\tilde\sigma,\textbf{R}^\prime \tilde\sigma) & = -\delta_{\textbf{R}\textbf{R}^\prime}\left[\left(M - \frac{4B}{a^2}\right)\tilde\sigma + \frac{4D}{a^2}\right] - \left(\frac{4B}{a^2}\tilde\sigma - \frac{4D}{a^2}\right)\left(\delta_{\textbf{R+a}_1,\textbf{R}^\prime}+\delta_{\textbf{R-a}_1,\textbf{R}^\prime}+\delta_{\textbf{R+a}_2,\textbf{R}^\prime}+\delta_{\textbf{R-a}_2,\textbf{R}^\prime}\right)\\
%\hat\epsilon_{2D,\sigma}(\textbf{R}a,\textbf{R}^\prime b) & = \frac{A}{2a}\left(\delta_{\textbf{R+a}_2,\textbf{R}^\prime}-\delta_{\textbf{R-a}_2,\textbf{R}^\prime}-i\delta_{\textbf{R+a}_1,\textbf{R}^\prime}+i\delta_{\textbf{R-a}_1,\textbf{R}^\prime}\right)\\\
%\hat\epsilon_{2D,\sigma}(\textbf{R}b,\textbf{R}^\prime a) & = -\hat\epsilon^*_{2D,\sigma}(\textbf{R}a,\textbf{R}^\prime b) \ .\\
%\end{align}
%\end{subequations}
\begin{subequations}
\begin{align}
\hat\epsilon_{2D,s}(\textbf{R}\tilde\sigma,\textbf{R}^\prime \tilde\sigma) & = \delta_{\textbf{R}\textbf{R}^\prime}\left[\left(M - \frac{4B}{a^2}\right)\tilde\sigma - \frac{4D}{a^2}\right] + \left(\frac{4B}{a^2}\tilde\sigma + \frac{4D}{a^2}\right)\left(\delta_{\textbf{R+a}_1,\textbf{R}^\prime}+ \delta_{\textbf{R-a}_1,\textbf{R}^\prime}+\delta_{\textbf{R+a}_2,\textbf{R}^\prime}+\delta_{\textbf{R-a}_2,\textbf{R}^\prime}\right)
\\
\hat\epsilon_{2D,\uparrow}(\textbf{R}a,\textbf{R}^\prime b) & = -\hat\epsilon_{2D,\downarrow}(\textbf{R}b,\textbf{R}^\prime a)  = \frac{A}{2a}\left(\delta_{\textbf{R+a}_2,\textbf{R}^\prime}-\delta_{\textbf{R-a}_2,\textbf{R}^\prime}-i\delta_{\textbf{R+a}_1,\textbf{R}^\prime}+i\delta_{\textbf{R-a}_1,\textbf{R}^\prime}\right)
\\
\hat\epsilon_{2D,\uparrow}(\textbf{R}b,\textbf{R}^\prime a) & = -\hat\epsilon_{2D,\downarrow}(\textbf{R}a,\textbf{R}^\prime b)  = -\hat\epsilon^*_{2D,\uparrow}(\textbf{R}a,\textbf{R}^\prime b) \, .
\end{align}
\end{subequations}
\end{widetext}

\end{document}